\documentclass[12pt]{iopart}

\usepackage{times}
\usepackage[OT1]{fontenc}
\usepackage[latin1]{inputenc}
\expandafter\let\csname equation*\endcsname\relax
  \expandafter\let\csname endequation*\endcsname\relax 
\usepackage{amsmath}
\usepackage{graphicx}
\usepackage{amssymb}
\usepackage{iopams}


\begin{document}

\newcommand{\ad}{a^{\dag}}
\newcommand{\rr}{{\bf r}}

\newcommand{\rC}{\text{\bf{C}}} 
\newcommand{\rI}{\text{\bf{I}}} 
\newcommand{\PC}{\mathcal{P}_{\rC}}
\newcommand{\bef}{\hat{\psi}}
\newcommand{\bfI}{\bef_{\rI}}
\newcommand{\cf}{\psi_{\rC}} 
\newcommand{\cfone}{\psi_{\rC1}}
\newcommand{\cftwo}{\psi_{\rC2}}
\newcommand{\cfp}{\psi_{\rC'}} 
\newcommand{\ecut}{\epsilon_{\rm cut}} 
\newcommand{\CF}{c-field}
\newcommand{\Nc}{N_{\rm{cond}}}
\newcommand{\psic}{\psi_{\rm{cond}}}
\newcommand{\ac}{\alpha_{\rm{cond}}}


\title{Temporal correlations of elongated Bose gases at finite temperature}
\author{Alice Bezett\footnote{ Current address: Institute for Theoretical Physics, Utrecht University,
Leuvenlaan 4, 3584 CE Utrecht, The Netherlands} and Emil Lundh}
\address{Department of Physics, Ume{\aa} University, SE-90187 Ume{\aa}, Sweden}
\begin{abstract}
Temporal correlations in the harmonically trapped finite temperature Bose gas are studied through the calculation of appropriate phase correlation functions. A wide parameter regime is covered to ascertain the role that temperature fluctuations and trap geometry play in the temporal coherence of the 1D to 3D crossover region. Bogoliubov analysis is used to establish results in the 1D and spherical limits. Formalism is then developed using the projected Gross-Pitaevskii equation to calculate correlation functions in 3D simulations of varying trap elongation and temperature. 
\end{abstract}

\maketitle
\section{Introduction}
\label{sec:introduction}
Macroscopic phase coherence is one of the key properties of a Bose-Einstein condensate (BEC), and has been a topic of longstanding interest. Almost as soon as BECs were created in the lab, this astounding property was demonstrated and investigated with respect to both spatial and temporal phase coherence \cite{ketterle1997, hall1998, greiner2002, orzel2001}. Understanding phase evolution is key to the development of applications such as interferometry using condensates, and a great deal of work has been done in this area, see for example \cite{shin2004,jo2007,esteve2008}. 
In addition, probing dynamic correlation functions by Bragg scattering is by now a mature technology \cite{griffin2003,papp2008}. 
While much progress has been made, fundamental questions still remain over the nature of phase fluctuations with regard to the role of geometry and finite-temperature interactions \cite{dettmer2001,hofferberth2007}.

The inclusion of a harmonic trapping potential adds complexity to the theoretical description of the gas, and the geometry of the trap dramatically alters the phase coherence properties \cite{dettmer2001,hofferberth2007,Scott2009}.
In three dimensions, phase coherence can be spatially extensive and long-lived, even at finite temperatures \cite{sinatra2009}; in a tightly confined trap within the one dimensional limit, however, long range coherence is lost over the length of the condensate, and quasicondensates of length shorter than the system form. In this regime, theory has been developed \cite{Demler_2007,Schmiedmayer_2009} to calculate the coherence time of the system, and has seen good agreement with experiment \cite{hofferberth2007}.

The cross-over regime between the 1D and 3D limits is significantly more complex at finite temperature, and formalism is required that can accurately describe the full-dimensional trapped cloud, and capture the interactions in the system exactly. One appropriate class of formalisms are the so-called c-field methods, which have already had much success in calculating correlations within the finite temperature BEC \cite{Scott2009,sinatra2009,wright2010}.

In this work, we study the transition region between the 1D and 3D limits. We make analytical calculations in the two extremes, and then develop formalism with the projected Gross Pitaevskii equation (PGPE) \cite{blakie2005,blakie2008} to numerically study the cross-over region and compare with our analytic results. 

This paper is organised as follows. We begin in Section \ref{sec:system} with our system and correlation definitions. Bogoliubov analysis for the quasi 1D and spherical limits is undertaken in Section \ref{sec:Bog}. In Section \ref{Formalism} we outline the PGPE formalism we will use to make numerical studies of the finite temperature Bose gas. In Section \ref{sec:results} we give the results of our simulations for a variety of system symmetries and temperatures. We conclude in Section \ref{sec:concl}.

\section{System and Correlation Functions}
\label{sec:system}

Our system is described by  the second quantized Hamiltonian
\begin{eqnarray}
\hat{H} &=& \int  d\mathbf{r}\, \hat\Psi^{\dagger}(\mathbf{r})\left\{ H_0   + \frac{U_0}{2}\hat\Psi^{\dagger}(\mathbf{r})\hat\Psi(\mathbf{r})\right\}\hat\Psi(\mathbf{r}) ,
\end{eqnarray}
where 
\begin{equation}
H_0=-\frac{\hbar^2}{2m}\nabla^2+V_0(\mathbf{r}),
\end{equation}
is the single particle Hamiltonian, $\hat\Psi(\mathbf{r})$ is the quantum Bose field operator, and $U_0 = 4\pi\hbar^2a/m$ is the interaction strength, with $a$ the s-wave scattering length. The trap potential is given as
\begin{equation}
V_0(\mathbf{r}) = \frac{1}{2}m\left(\omega_x^2x^2+\omega_y^2y^2+\omega_z^2z^2\right),
\end{equation}
where $\{\omega_x,\omega_y,\omega_z\}$ are the trap frequencies.

In this work we will study the phase correlations within a system with respect to time. The phase correlation function between two points $\mathbf{r_1},\mathbf{r_2}$ and two times $t_1,t_2$ is given as
\begin{equation}
g(\rr_1,t_1,\rr_2,t_2) = \langle \hat\Psi^\dagger(\rr_1,t_1)\hat\Psi(\rr_2,t_2)\rangle \label{g}, 
\end{equation}
where the average is taken over realizations of field configurations. We also define a spatially averaged autocorrelation function, 
\begin{equation}
g_{11}(t_1,t_2) = \sum_{\rr} g(\rr,t_1,\rr,t_2) \label{g11},
\end{equation}
where the points $\rr$ are chosen to lie in the interior of the cloud where the density is high. This autocorrelation function gives the phase correlation of a single point in the system with respect to time, and we indicate this with the subscript '$11$'. We also define the two point autocorrelation function
\begin{equation}
g_{12}(t_1,t_2) = \sum_{<\rr_1,\rr_2>} g(\rr_1,t_1,\rr_2,t_2) \label{g12},
\end{equation}
where both $\rr_1$ and $\rr_2$ are taken at interior points of system 1 and system 2, where the respective densities are high. This intercloud autocorrelation shows the phase correlation between two different systems with respect to time, and we signify this with the subscript '$12$'.

We also define the normalised versions of Eqs. (\ref{g},\ref{g11},\ref{g12}) as
\begin{eqnarray}
G(\rr_1,t_1,\rr_2,t_2) &=& \frac{g(\rr_1,t_1,\rr_2,t_2)}{\sqrt{g(\rr_1,t_1,\rr_1,t_2) g(\rr_2,t_1,\rr_2,t_2)}}\\
G_{11} &=& \frac{\sum_{i=1}^n G(\rr_i,t_1,\rr_i,t_2)}{n}\\
G_{12} &=& \frac{\sum_{i,j=1}^n G(\rr_i,t_1,\rr_j,t_2)}{n}
\end{eqnarray}

\section{Bogoliubov Analysis}
\label{sec:Bog}

In this section, we solve the Bogoliubov equations to find the decay of the autocorrelation within a cloud with respect to time. This is done in the quasi-1D and spherical limits. These analytical results will form the basis from which we will compare our numerical results.

\subsection{Quasi-1D Systems}

For a very elongated gas, the trap potential may be written 
\begin{equation}
V_{el}(x,y,z) = \frac12 m\left[\omega_{\perp}^2(x^2+y^2) + \omega_z^2 z^2\right].
\end{equation}
For later use, we define the asymmetry parameter 
\begin{equation}
\lambda=\frac{\omega_z}{\omega_{\perp}},
\end{equation}
and the oscillator lengths 
$a_{\perp}=(\hbar/m\omega_{\perp})^{1/2}$ and 
$a_z = a_{\perp}/\lambda^{1/2}$.
We assume first $\lambda \ll 1$, so that the excitation of normal 
modes in the perpendicular direction can be neglected. For the case we consider, it is assumed that the interaction energy is high enough that the condensate occupies 
several trap modes, which enables us in the analytical calculations to treat 
the condensate wave function in the Thomas-Fermi approximation 
\cite{pethick2008book}, 
\begin{equation}
n(x,y,z) = \frac{\mu}{g} \left(1-\frac{x^2+y^2}{R^2} - \frac{z^2}{Z^2}\right),
\end{equation}
where 
\begin{eqnarray}
R^2 &=& \frac{2\mu}{m\omega_{\perp}^2},\nonumber\\
Z^2 &=& \frac{2\mu}{m\omega_{z}^2},
\end{eqnarray}
and finally 
\begin{equation}
\mu = \frac{\hbar \omega_z}{2} \left(\frac{15 N a}{\lambda^2 a_z}\right)^{2/5}.
\end{equation}

The calculation of autocorrelation functions for a Bose gas in the Bogoliubov 
approximation was outlined in Refs.\ \cite{griffin2003,petrov2000,cetoli2010}. 
The correlation function is written 
\begin{equation}
g_{11}(\rr,t,\rr',0) = \sqrt{n(\rr) n(\rr')} \exp\left[-\frac12 F(\rr,t,\rr',0)\right],
\end{equation}
where 
\begin{equation}
F(\rr,t,\rr',0) = \langle |\hat{\phi}(\rr,t) - \hat{\phi}(\rr',0)|^2\rangle,
\end{equation}
and the phase operator is defined as 
\begin{equation}
\hat{\phi}(\rr,t) = \sum_j \left[\frac{f_j^{-}(\rr)}{\sqrt{n(\rr)}}e^{i\omega_j t} \hat{a}_j + {\rm h.c.} \right],
\end{equation}
where finally $f_j^{-}(\rr)$ is a solution to the 
Bogoliubov equations in number-phase form \cite{fetter1996}, 
\begin{eqnarray}
\left[L-gn(\rr)\right] f_j^{-}(\rr) &=& \omega_j f_j^{+}(\rr),
\nonumber\\
\left[L+gn(\rr)\right] f_j^{+}(\rr) &=& \omega_j f_j^{-}(\rr),
\end{eqnarray}
and the linear operator $L$ is defined as 
\begin{equation}
L = -\frac{\hbar^2}{2m} \nabla^2 + V(\rr) -\mu+2gn(\rr).
\end{equation}
The number-phase representation functions $f_j^{\pm}$ used here are 
related to the more familiar 
Bogoliubov amplitudes $u_j$ and $v_j$ through 
$f_j^{\pm}(\rr) = [u_j(\rr)\pm v_j(\rr)]/2$, where we have used the sign and phase 
conventions of Ref.\ \cite{pethick2008book}. 
The solution for the longitudinal modes, using the Thomas-Fermi approximation 
for the condensate density, is \cite{petrov2000,stringari1997} 
\begin{eqnarray}
f_j^{-}(z) &=& \left[\frac{2gn(\rr)}{\omega_j}
\frac{1}{8\pi R^2 Z}\frac{(j+2)(2j+3)}{j+1}\right]^{1/2}
P_j^{(1,1)}(z/Z), \nonumber\\
\omega_j &=& \omega_z \sqrt{\frac{j(j+3)}{4}},
\label{bogoliubovfreq}
\end{eqnarray}
where $j=1,2,\ldots$, and  $P_{j}^{(a,b)}(x)$ is a Jacobi polynomial 
\cite{abramowitz1965book}. 
Computing the temporal correlation at the origin, $\rr=\rr'=0$, for a 
temperature $T$ assumed to be much larger than the longitudinal trap 
frequency $\omega_z$,
we obtain 
\begin{eqnarray}
F(0,t,0,t) &=& \frac{g}{\pi R^L}\sum_{j \, \rm even} 
\frac{(j+2)(2j+3)}{j+1}\frac{1}{\omega_j}\frac{T}{\omega_j} \nonumber\\
&&\times\left(\frac{2(j+1)!!}{(j+2)!!}\right)^2(1-\cos \omega_j t).
\end{eqnarray}
It is found that at short times, $t \ll \omega_z^{-1}$, $F$ is proportional to the 
square of $t$, i.e. $g_{11}\propto e^{-t^2}$. 
For moderately short times, $t \sim \omega_z^{-1}$, $F$ goes linear in $t$ and 
the result for the correlation function is 
\begin{equation}
g(0,t,0,t) \propto e^{-\Gamma \omega_z t}, 
\end{equation}
where 
\begin{equation}
\Gamma = \frac{16\pi}{15}\frac{k_B T}{\hbar\omega_z} 
\left(\frac{15 Na}{\lambda^2 a_z}\right)^{2/5}.
\end{equation}
For times $t \gtrsim \omega_z^{-1}$, the correlation function will 
exhibit a quasiperiodic oscillatory behaviour.

\subsection{3D Systems}

To contrast, we now treat analytically the case of a spherical 
condensate, with $\omega_{\perp}=\omega_z=\omega$. 
The excitation spectrum is given by
\cite{stringari1997} 
\begin{eqnarray}
f_n^{-}(r) = \left(\frac{4n+3}{\pi R^3}\frac{2g}{\omega_n}\right)^{1/2} 
P_n^{(0,1/2)}(2r^2/R^2-1), 
\nonumber\\
\omega_n = \omega n (2n+3),
\end{eqnarray}
where $n$ are positive integers. 
Following the method in the previous section, the exponent of the correlation function then becomes 
\begin{eqnarray}
F(0,t,0,t) &=& \frac{4k_BT}{\hbar\omega} \frac{2g}{\pi R^3 \omega}
\sum_{n=1}^{\infty} \frac{4n+3}{[n(2n+3)]^2} \nonumber\\
&&\times\left[\frac{(2n+1)!}{2^{2n} (n!)^2}\right]^2 (1-\cos\omega_n t).
\end{eqnarray}
At very short times, $t\ll\omega_z^{-1}$, F is once again proportional to the square of t. However, at short times $t \sim \omega_z^{-1}$ F is approximately equal to 
\begin{equation}
F(0,t,0,t) = \frac{8 g k_BT}{R^3 \hbar^2\omega^2}\sqrt{\frac{\omega t}{\pi}},
\end{equation}
so in the spherical case, the correlation goes as the square root of $t$ for 
short times, and this distinction will be used in later sections to make comparisons between different trap symmetries. As in the previous case, when $t \gtrsim \omega^{-1}$, quasiperiodic behavior 
ensues. 

\section{PGPE Formalism}
\label{Formalism}

Motivated by the need to develop tractable theory to describe experimentally realistic systems,
we now move to a system intermediary to the quasi 1D and spherical 3D systems, that is to say, the elongated Bose gas. We develop the PGPE method to calculate temporal phase correlations in a three dimensional Bose gas at finite temperature. In this particular work, we note that it has a key advantage over Bogoliubov methods: the ease with which this method can accurately calculate temporal correlations for a wide variety of physical applications. We demonstrate this later in Section \ref{sec:results} by calculating results for differing trap aspect ratios and temperatures.

In what follows, we give an overview of the PGPE method, and our sampling method for studying data. More details on the PGPE and c-field methods can be found in Refs.\ \cite{blakie2005,blakie2008,blakie_review}.

\subsection{PGPE Formalism and Correlation Functions}

The key aspect of this formalism is that the Bose field operator is separated into two parts, dependent on the mean occupation of modes in the system. This splitting is written as
\begin{equation}
\hat\Psi(\mathbf{r}) = \cf(\mathbf{r}) + \bfI(\mathbf{r}),\label{EqfieldOp}
\end{equation}
where $\cf$ is the coherent region \CF\  and $\bfI$ is the incoherent field operator \cite{blakie2005}.
These fields are defined as the low and high energy projections of the full quantum field operator, separated by the energy $\ecut$
. In our theory this cutoff is implemented in terms of the harmonic oscillator eigenstates $\{\varphi_n(\mathbf{r})\}$ of the time-independent single particle Hamiltonian
i.e. $\epsilon_n\varphi_n(\mathbf{r})=H_0\varphi_n(\mathbf{r})$, with $\epsilon_n$ the respective eigenvalues.

In general, the applicability of the PGPE approach to describing the finite temperature gas relies on an appropriate choice for $\ecut$, so that the modes at the cutoff have an average occupation of order unity. This choice means that all the modes in $\rC$ are appreciably occupied, justifying the use of a classical field for these modes. In contrast the $\rI$ region contains many sparsely occupied modes that are particle-like and would be poorly described using a classical field approximation.

The equation of motion for $\cf$ is the PGPE
\begin{eqnarray}
i\hbar\frac{\partial \cf }{\partial t} = H_0\cf + \PC\left\{ U_0 |\cf|^2\cf\right\}, \label{PGPE}
\end{eqnarray}
where the projection operator 
\begin{equation}
\PC\{ F(\mathbf{r})\}\equiv\sum_{n\in\rC}\varphi_{n}(\mathbf{r})\int
d\mathbf{r}'\,\varphi_{n}^{*}(\mathbf{r}') F(\mathbf{r}'),\label{eq:projectorC}\\
\end{equation}
formalises our basis set restriction of $\cf$ to the $\rC$ region. The main approximation used to arrive at the PGPE is to neglect dynamical couplings to the incoherent region \cite{Davis2001b}. 
The $\rC$ and $\rI$ regions are then treated as independent systems in thermal and diffusive equilibrium.

An important feature of Eq. (\ref{PGPE}) is that it is ergodic, so that  the microstates the classical field $\cf$ evolves through in time form a sample of the equilibrium microstates, and  time-averaging can be used to obtain macroscopic equilibrium properties.

The procedure for finding equilibrium states including detail on using appropriate input states is given elsewhere \cite{Bezett2009}. Here, we focus on the method necessary for studying time correlations in the system. We define the temporal correlation function for the coherent region as
\begin{equation}
g_{\bf{C}}(t_1,t_2) = \langle \psi^*_{\bf{C}}(\mathbf{r},t_1) \psi_{\bf{C}}(\mathbf{r},t_2) \rangle \label{g1_classical},
\end{equation}
i.e. we replace the full Bose field operator with the coherent region c-field. Analogues of $G_{12}$ and $G_{11}$ follow from this, as previously.

In previous studies \cite{Bezett2008} it was shown that the contribution to the spatial phase correlation function by the incoherent region was negligible, as any contribution was due only to thermal atoms and existed over very short range. Here we neglect the incoherent region when calculating phase correlations. Contribution from purely thermal atoms are of course already in the coherent region, via the higher energy modes included there.

We approximate the average in Eq. \ref{g1_classical} with an ensemble average. Our procedure for calculating phase correlations within the PGPE is as follows:
\begin{itemize}
\item Create an ensemble of 40 input states each with different random fluctuations, and evolve these within the PGPE
\item At equilibrium, calculate $\psi^*_\mathbf{C}(\mathbf{r},0) \psi_\mathbf{C}(\mathbf{r},t)$ for each of the 40 simulations
\item Perform the average over the ensemble of 40 simulations to calculate $\langle \psi^*_\mathbf{C}(\mathbf{r},0) \psi_\mathbf{C}(\mathbf{r},t) \rangle$.
\end{itemize}

\section{PGPE Results and Analysis}
\label{sec:results}

In this section we present the results of the application of our PGPE formalism to a cold Bose gas. We consider a system of $^{23}$Na atoms in the elongated traps with frequencies $\{ \omega_x,\omega_y,\omega_z\} = 28\pi\{10,10,1\}$Hz and $\{ \omega_x,\omega_y,\omega_z\} = 28\pi\{20,20,1\}$Hz . We study two temperatures at each trap, corresponding to condensate fractions 45$\%$ ($T/T_c\approx0.62$) and 10$\%$ ($T/T_c\approx0.8$). As the condensate is responsible for any long-time coherence in the system, we keep the condensate number similar between the two trap symmetries, with $N_{cond}\approx 45000$ for the cooler system, and $N_{cond}\approx30000$ for the hotter. 

\subsection{Density at Equilibrium}

\begin{figure}
\includegraphics[width=3in,keepaspectratio]{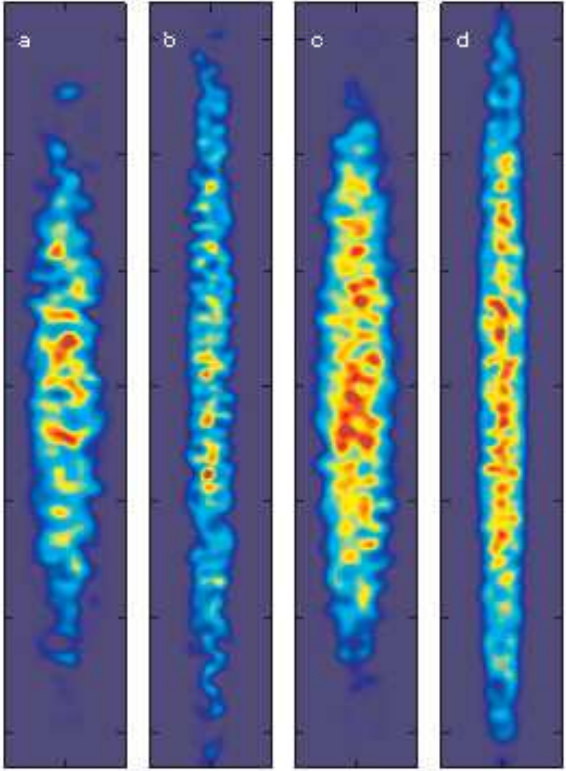} 
\caption{Density slice in the y=0 plane for the four systems studied. Each subfigure is normalised separately. From left to right, a) Aspect ratio $\lambda=10$, $T/T_c\approx0.8$, b) $\lambda=20$, $T/T_c\approx0.8$, c) $\lambda=10$, $T/T_c\approx0.62$, d) $\lambda=20$, $T/T_c\approx0.62$}\label{density}
\end{figure}

We begin with a general discussion of the properties of the densities of our four different systems. A density slice in the $y=0$ plane is shown for each in Fig \ref{density} (see caption).

The density graphs show one instance of the distribution as it evolves in time. The three dimensional PGPE is ergodic, and as such, each simulation evolves through the set of allowable microstates. In this way, a single shot from a PGPE simulation can be likened to a single shot from an experiment - i.e., it gives a microstate rather than an equilibrium average. This dynamic quality can easily be seen in the four microstates shown in Fig. \ref{density}, where the density for each is highly irregular.
As well as showing the effect of the changing aspect ratio on the basic shape of the cloud, these slices also show the increasing fragmentisation of the condensate density at high temperatures or aspect ratios \cite{dettmer2001}. This is perhaps most striking in Fig \ref{density}b, where high temperature and high aspect ratio cause the density to form discrete high density clumps, as seen in the separated circular features approximately halfway down the figure.

\subsection{Autocorrelation}

\begin{figure}
\includegraphics[width=3.5in,keepaspectratio]{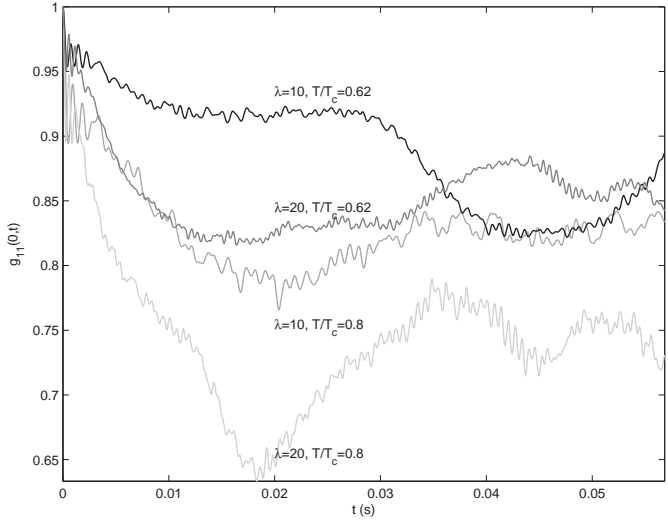} 
\caption{Autocorrelation function $g_{11}$ for four systems of study: variables given in Figure.}\label{auto_fig}
\end{figure}

The autocorrelation is calculated according to Eq. (\ref{g1_classical}), and the results are presented in Fig \ref{auto_fig} for our four systems (see caption).

The autocorrelation for all cases comprises of three distinct parts: an initial sharp decay, subsequent larger decay, and long-term quasiperiodic behaviour.

The initial sharp decay occurs on a timescale $t\ll\omega_z^{-1}$ for all systems, and represents the initial loss of coherence over short timescales by the thermal modes of the system. This is due to the significant proportion of the thermal cloud included in the PGPE description.

The subsequent decay in autocorrelation is due to the changing phase of the condensate in each system with respect to time, and it is this feature of the autocorrelation that differs so markedly between each system. It can clearly be seen that the proportion of condensate present in a system affects how quickly it will decohere, as can be seen by comparing curves of different temperature and same trap (see caption). Perhaps more surprising is the marked difference that the aspect ratio of the trap makes to the decay in coherence. For simulations with similar condensate fraction and number, but different aspect ratios, the tighter trap leads to a much more rapid decoherence, as can be seen by comparing simulations of similar temperature and different aspect ratio. For the two $T/T_c\approx0.62$ systems (Figure (2), top two curves) the $\lambda=10$ system has $g_{11}=0.92$ at $t=0.02$s, while the $\lambda=20$ system has $g_{11}=0.82$. This is a large difference for the very modest changes in aspect ratio we simulate. This more rapid phase decoherence is due to the lack of a spatially extensive condensate in the $\lambda=20$ case causing greater phase fluctuation along the system length.

Following this decay, there is quasiperiodic behaviour in the system, shown as a period of revival and further decays of the autocorrelation function, on the timescale $t>\omega^{-1}_z$. This is consistent with the prediction from our Bogoliubov anaylsis results.

There can also be seen on both curves a fine structure noise. This occurs at a frequency of $\omega_x$, and is due to fluctuations in the tight dimensions of the trap. We note that the PGPE description includes the effects and interactions of all appreciably occupied modes, and for these systems, there are several modes in the tight direction that are occupied. Since we are interested only in coherence along the long axis, we made no further analysis of these fluctuations.

In order to make a quantitative comparison of our results with those of Section \ref{sec:Bog}, we plot $\textrm{ln}(-\textrm{ln}(g_{12}))$ against $\textrm{ln}(t)$, and extract the gradient from the line of best fit. Results are given in Table \ref{tab:allfits}, both for very short times $t\ll\omega^{-1}_z$ and short times $t\approx\omega^{-1}_z$. Comparing our results, we see that at very short times, all results correspond approximately to $2$, i.e. $g_{11}\propto e^{-t^2}$ as expected. For $t\approx\omega_z^{-1}$, however, it is found that the exponent is different for each case. From Bogoliubov analysis, the expected result for the exponent of the decay is $0.5$ for the 3D case, and $1$ for the 1D case. The results for aspect ratio $\lambda=10$ trap confirm that this system is still well within the 3D regime. For aspect ratio $\lambda=20$ however, it can be seen that the results are intermediary to the two predictions. This can be explained in terms of the relative temperature to tight aspect ratio in our simulations. For aspect ratio 10, the temperature required for excitations in the tight trap direction is $T\approx \frac{\hbar \omega_x}{k_B} = 6.7$nK, and the temperature of the cooler system is ten times this, and shows clear excitation of the axial modes. With an aspect ratio 20, $T\approx \frac{\hbar \omega_x}{k_B} = 13$nK, and the simulation presented here has a temperature 8 times this value. In this case, it can be seen that more of the axial modes are 'frozen out', and the fitting of the exponent gives a result intermediary to the 1D and 3D limits, as expected.

\begin{table}[htbp]
\centering
\label{tab:allfits}
\begin{tabular}{ |c |c|cc|}
	\hline
$\mathbf{\lambda}$  & $T/T_c$ &  $t<\omega_z^{-1}$ & $t\approx\omega_z^{-1}$      \\
	\hline \hline
10 &  0.62 &2.0 & 0.49   \\
20 & 0.62 & 1.9 & 0.71  \\
10 & 0.8 & 2.0 & 0.44  \\
20 & 0.8 & 1.9 & 0.62   \\ 
	\hline \hline
\end{tabular}
\caption{Results of fit to autocorrelation function $g_{11}$, for four systems, and two time limits.}
\label{tab:allfits}
\end{table}

It can also be seen in the fitted results that the exponent at times $t\approx\omega_z^{-1}$ decreases with increasing temperature, leading to a faster decay in $g_{11}$. This is expected as the condensate fraction is responsible for long time coherence, and this is diminished at higher temperatures.

\begin{figure}
\includegraphics[width=3.5in,keepaspectratio]{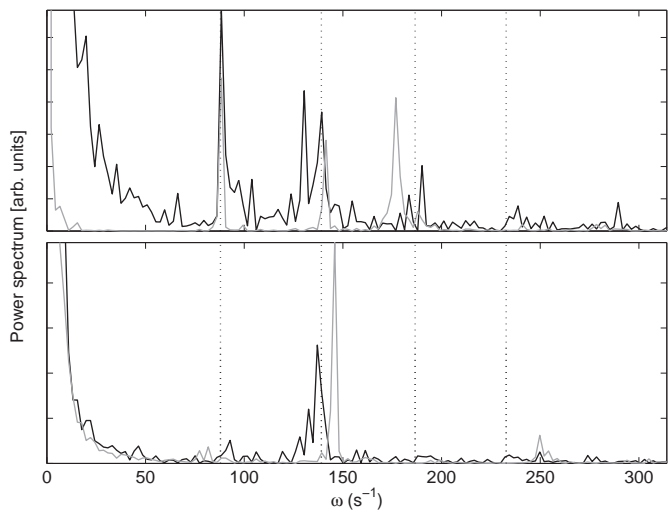}
\caption{Results of fourier analysis. Top: aspect ratio $\lambda=10$, $T/T_c\approx0.8$ (black), $T/T_c\approx0.55$ (grey), prediction from 1D Bogoliubov analysis (dotted). Bottom: $\lambda=20$, $T/T_c\approx0.8$ (black), $T/T_c\approx0.55$ (grey),prediction from 1D Bogoliubov analysis (dotted). }
\label{fourier}
\end{figure}

As a further means of analysis, we calculate the frequency spectrum of the system from the autocorrelation function as
\begin{equation}
F_{\psi_C^*\psi_C}(\omega) = \int \int_{-\infty}^{-\infty}d\mathbf{r}dte^{i\omega t} \langle\psi_C^*(\mathbf{r},t)\psi_C(\mathbf{r},0)\rangle.
\end{equation}
The corresponding power spectra for the four systems are shown in Figure \ref{fourier}. For $\lambda=10$, at low temperature we see a set of well defined peaks close to the 1D Bogoliubov frequencies given in Eq.\ (\ref{bogoliubovfreq}). At higher temperature a number of features is observed: Broadening of peaks, shifting of frequencies, possibly splitting of modes and what appears to be a few new peaks that do not correspond to a 1D Bogoliubov frequency. 
For $\lambda=20$, 
one mode appears to dominate, corresponding to the mode with quantum number $n=2$. This corresponds to the longitudinal breathing mode of the system, as would be expected.

\section{Conclusions}
\label{sec:concl}

The finite temperature Bose gas is a nontrivial system displaying complex properties of phase coherence. The transition between coherence in the 1D and 3D limits has been little studied, and here we seek to determine the system behaviour in this regime. We develop formalism based on the PGPE to numerically study these symmetries, and compare this to our results from Bogoliubov analysis of the 1D and spherical limits. It is found that systems with aspect ratio 10 and 20 both give exponents for the decay of the correlation time that are intermediate to the 1D and 3D limits. This can be explained in terms of the finite temperature of the system causing excitations to be present in the tight directions of the trap, placing our systems of study deeply within the 1D to 3D crossover region.

For experimental studies of the phase evolution of the systems studied here, a splitting and recombination method \cite{hofferberth2007} would be impractical, since with reasonably high temperatures, splitting and recombining in real space is too slow to capture the initial decay of phase coherence. 
In this case the systems would be better probed with a Bragg-type measurement \cite{griffin2003,papp2008} in which atoms are outcoupled using a light pulse, and the momentum-dependent power spectrum is directly obtained.

This project was financially supported by the Swedish Research Council, 
Vetenskapsr{\aa}det, and the Kempe foundation. This research was conducted using the resources of the High Performance 
Computing Center North (HPC2N).


\end{document}